# Prediction of a Mobile Solid State in Dense Hydrogen under High Pressures

Hua Y. Geng[*], Q. Wu, and Y. Sun

*National Key Laboratory of Shock Wave and Detonation Physics, Institute of Fluid Physics, CAEP; P.O.Box 919-102 Mianyang, Sichuan, P. R. China, 621900*

## Abstract

**Solid rigidity and liquid-scale mobility are thought incompatible in elemental substances. One cannot have an elemental solid that is long-range positionally ordered wherein the atoms flow like in a liquid simultaneously. The only exception might be the hypothetical supersolid state of $^4$He. In this work, we demonstrate that such exotic state could exist even in the classical regime. Using *ab initio* molecular dynamics (AIMD) and *ab initio* path integral molecular dynamics (AI-PIMD), a novel state of dense hydrogen which simultaneously has both long-range spatial ordering and liquid-scale atomic mobility is discovered at 1~1.5 TPa (1 TPa≈10000000 atmospheric pressures). The features distinct from a normal solid and liquid are carefully characterized, with the stability and melting behavior are investigated. Extensive AI-PIMD simulations further revealed that this state might be (meta-)stable even at ultra-low temperatures, suggesting an emerging candidate for an alternative type of supersolid state in dense metallic hydrogen.**

Table of Contents (TOC) Graphic

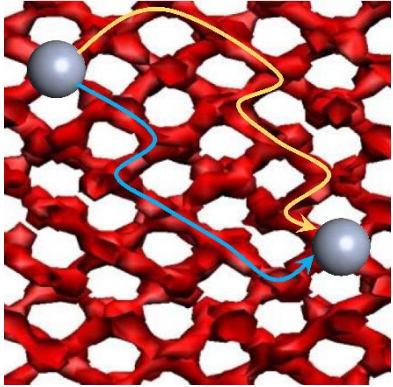

The solid state of matter is usually characterized by the structural rigidity and spontaneous resistance to the change in its shape against external forces. The particles in a typical solid are tightly bound to each other, and vibrate around their equilibrium sites. In a crystalline state, particles distribute in a regular geometric lattice, thus giving long-range positional ordering.[1] Amorphous or glassy states are also considered as solid, except there particles distribute irregularly. Usually a crystalline state must be a solid, and to crystallize is always equivalent to make a solid. The liquid state, on the other hand, is flowing, thus being homogeneous and isotropic. It is absence of any long-range spatial ordering, which is a consequence of the free mobility of particles at both microscopic and macroscopic scales in this state.[2,3] For this reason, anisotropy could also be considered as a precursor of crystallization. The intermediate state between them, or *mesomorphic* phase, is fascinating and attracts broad attention. One of the well-known examples is liquid crystal, which always involves highly anisotropic molecules or polymers.[3-5] Another category close to this concept is superionic compounds, in which the heavy ions form a solid framework whereas the light component becomes molten and migrates through the lattice as in a fluid.[6] Typical superionic solid has at least two or more constituents. Recently a similar phenomenon was also predicted in hydrides, where the hydrogen sublattice exhibits diffusive or liquidlike properties.[7,8] So far, there is no report on elementary mesophase in which the same element participates in both the crystalline framework and moving through the lattice, thus as a genuine *mobile solid*. In quantum regime, it has long been proposed that helium might have such an exotic state that was termed as supersolid, though yet to be confirmed by experiment.[9,10] In supersolid, large zero-point motion delocalizes self-vacancies of the matter to enter a Bose-Einstein condensate state, which allows $^4$He to become a superfluid while maintaining the solid lattice simultaneously, mainly due to the non-locality of quantum mechanics.[10] Note that the phase stability of crystalline helium increases rapidly with compression, thus it does not expect to have large atomic mobility at low temperature under high pressures.

Search for an elementary mesomorphic phase is a long-standing open issue in physics and chemistry. The difficulty mainly roots in the tremendous degrees of freedom in the many-body system, which leads to complicated correlations and even becomes intractable in theoretical analysis. In terms of the potential energy surface, a crystalline state may be represented by a point on that surface defined in the configurational space. A homogenous liquid, on the other hand, covers a broad region and comprises transient configurations that might be visited by the liquid state.[2] The exact characteristics of a mobile solid, however, are very hard to imagine. Its atomic motion pattern, which should maintain lattice geometry and undergo a large-scale liquid-like diffusion simultaneously, is quite counterintuitive.

Fast evolved advanced computer simulation provides an alternative route to investigate this challenging problem. By employing path integral and *ab initio* molecular dynamics (AIMD), we will demonstrate below for the first time that even in the classical regime the fluid-scale atomic diffusion and three-dimensional long-range positional ordering can coexist harmonically in elementary substance, with hydrogen under extreme compression as a probable example. In this bizarre state the same elemental atoms participate in both the crystalline lattice formation and the mass transportation (or hopping) between lattice sites. The predicted existence of a classical mobile solid demonstrates the richness of physics and chemistry in mesophases, and might eventually lead to an alternative realization of supersolid state other than helium.[10]

The calculation is carried out using AIMD (with protons being treated as classical) and path integral AIMD (AI-PIMD), in which density functional theory (DFT) is employed to describe the electronic structures within Born-Oppenheimer approximation. Details of the computation are given in Methods section. At pressures between 1 and 1.5 TPa, we frequently observed an interesting phase of liquid dense hydrogen at 50 K in AI-PIMD simulations. As shown in Fig. 1, in this phase the diffusivity of protons, as manifested as the root of the mean square displacement (rMSD) evaluated with AI-PIMD, is unusually high. It is orders higher by comparison with regular solids with defects or dislocations at similar temperatures.[11-13] By carefully inspecting the atomic structure and MD trajectories, we failed to find any well-defined dislocations or grain boundaries. Instead, it looks like all atoms participate in the diffusion process concertedly, though not simultaneously. Thus it is a bulk behavior, rather than migration confined to dislocations or boundaries. Because we know that bulk flowing is an intrinsic characteristic of liquid, this raises the speculation that in this phase the protons could be considered as a flow as in a liquid.

This observation is consistent with previous simulations, where it was taken as a normal liquid.[14] However, we noticed that though the calculated radial distribution function (RDF) gives a feature very similar to a typical liquid, a distorted lattice pattern still presents through the long-time averaged proton density distribution (the inset in the left panel of Fig. 1). This is at odds with the assumption that it is in an isotropic liquid state. Furthermore, the projected RDFs[15] (right panel in Fig. 1) unveil a strong anisotropy and long-range positional ordering. This is quite different from our understanding about a typical liquid, in which the particle density distribution must not have any local or long-range features and the system must be isotropic. The unexpected emergence of anisotropy implies that the matter still have some crystalline features, which is a strong indicator for a solid state. For these reasons, it seems plausible to tentatively interpret it as a novel matter state in which the solid and liquid features coexist harmonically in one pure phase.

On the other hand, the observed structure pattern and anisotropy are maintained over a wide temperature range. As shown in Fig. 1 that was calculated with AI-PIMD simulations at 100 K, the increase of temperature does not change the structure pattern very much, and the accumulated RDF and projected RDFs are almost the same as that at 50 K. But interestingly, the rMSD (thus the self-diffusion mobility) is greatly enhanced. In other words, in this phase the atomic diffusivity can be enhanced without at the expense of the *solid* framework.

This phenomenon inspires the conjecture that a genuine mobile solid could exist in this simple elemental substance. Alternatively, one might attempt to interpret this unexpected phenomenon by using the non-locality of quantum mechanics, being analogous to the hypothesized concept of supersolid.[10] Nonetheless, by closely inspecting the bead trajectories in AI-PIMD simulations, we found that the particle wave function is well localized, and no tunneling events were observed.[15] This signals that the quantum non-locality is not a necessity for this novel state to exist.

Based on these observations, we suppose that some specific aspects in the potential energy surface might be the key for this bizarre state to appear. If it is true, then the same phase should manifest even in the classical regime. In order to verify this supposition, AIMD simulations with protons being treated as classical were intentionally carried out, in which the electronic part was accurately described with DFT.

As envisioned above, the main features of the exotic state observed in AI-PIMD

simulations are perfectly preserved in classical AIMD simulations. Furthermore, the positional ordering becomes sharper and more distinct in the latter case. This could be due to that we have neglected the zero point motion here, which is large in hydrogen and always prefers to isotropic state. As shown in Fig. 2 (that marked as the fine potential curve), at low temperature limit, the diffusion constant is close to that of a typical solid state, and is at the same level of the migration of hydrogen impurity in solid iron.[16] However, the magnitude of the proton self-diffusion constant jumps about 4 orders higher when temperature increases from 100 to 300 K. It even becomes comparable to that of a typical liquid when above 225 K, as the hatched region drawn in Fig. 2 shows. Especially, the mobility of protons in this novel state is even comparable with that of air or hydrogen gas in water.[16] The rapid increase of the diffusivity might imply a phase transition into a normal liquid phase. Nonetheless, simulations unveiled that the averaged positional ordering is preserved very well even when temperature approaches 400 K, where the diffusion constant enters a new regime and marks the end of the transition (see the proton density distribution patterns illustrated in Fig. 2). Therefore it is not a conventional melting we are familiar with. Rather it can be compared to the liquid crystals of anisotropic molecules.

It is necessary to point out that nuclear quantum effect (NQE) usually broadens particle distribution. This is analogous to the classical kinetic effects in solid or liquid.[17,18] Therefore a system with large quantum dispersion at low temperature can *qualitatively* be mirrored to the classical counterpart at higher temperatures if considered only the long-time averaged atomic structural features.[15,19] We thus speculate that the distortion or imperfectness presented in the structure as shown in Fig. 1 could be mainly due to NQE. This argument is corroborated by Fig. 2, in which a similar proton density distribution pattern is found. But the distinct crystalline features gradually fade away with increased classical kinetic motion of particles.

On the other hand, even in these AIMD simulations the jump in the diffusivity is not very sharp. It proceeds with several stages. This might originate in the complicated energy landscape of dense hydrogen around this pressure range.[20,21] In order to sharpen our understanding and to acquire a more clear physics picture, it might be helpful to go beyond the true physical world and use less accurate interactions that can capture the main characteristics of the potential surface. This can be done by a slightly coarse sampling of the *k*-points in DFT calculations. Degrading the *k*-point sampling resolution from a fine grid of 2×2×2 to two special *k*-points (the 2KPM as defined in Ref. 15) keeps the essential physics observed here. Note that the resultant coarse effective interatomic potential does not correspond to the real dense hydrogen accurately. Rather it is a physical model to aid us in understanding about the underlying mechanism of the novel state just discovered.

In this approximate model, the transition from the normal solid to the exotic state becomes clean-cut, as exhibited in Figs. 2 (the coarse potential curve) and 3. The corresponding normal solid at low temperatures is a perfect crystalline phase. It contains no defects or grain boundaries. The primitive cell has 48H with a space group of $P2_1/c$. The transition to the exotic state at ~350 K greatly enhances the proton mobility, which is parallel to the results of the fine potential calculated with accurate DFT simulations. The positional ordering (as unveiled by the proton density distributions in Fig. 2) is perfectly preserved after the transition finished, unequivocally demonstrates that it is not a typical liquid. It is worthwhile to remark that the increased diffusivity cannot be due to defect migration, since the latter is of orders smaller. For example, hydrogen diffusion constant is only about $10^{-8}$~$10^{-7}$ Å$^2$/fs at 672~823 K in solid LiH with point defects.[22]

In this model system, the calculated classical rMSD exhibits a typical liquid diffusivity at temperatures above 350 K. But at low temperatures it is of typical solid characteristics [Fig. 3(c)]. This is in good agreement with the diffusion constant shown in Fig. 2. Please notice the small hysteresis in Fig. 3(b) that occurs at 340 and 350 K, which suggests that the transition from *P2$_1$/c* is first order.

Besides the diffusivity and rMSD, a sharp change in RDF $g(r)$ also can be found at about 350 K [Fig. 3(a)]. Beyond this transition, the second peak of $g(r)$ becomes diminishing. Therefore, if inspecting only the short-range correlations, say, less than 2 Å, one might incorrectly conclude it as a conventional melting to the isotropic liquid.[14] Nonetheless, long-range correlations actually persist up to 970 K (see also Fig. S4), and the true melting temperature of this model system thus must be higher.

An intuitive comprehension about this state can be obtained via the proton density distribution from MD simulations, as well as its correlation with pressure/potential energy. For a normal solid, the particle density distribution usually forms well localized and separated pockets even at the vicinity of the melting temperature, as illustrated by the gold as an example in the Fig. S3 of the Supporting Information. For dense hydrogen in the 2KPM model approximation, however, there is a sharp jump in both the pressure and potential energy at ~350 K along the isochore that manifested in Fig. 3(b), which is in accordance with the exotic state transition as presented in Figs. 1 and 2. On the other hand, the proton density distribution forms isolated pockets when below 350 K, showing that under this condition the model system is a typical solid. Nevertheless, as the temperature increases, these pockets develop into interconnections and form diffusion channels. This unveils the collective nature of proton drift in this state. In other words, it is the long-time averaged diffusion channels rather than the transient coordinates of individual particles that constitute the observed *lattice*. This mechanism is simply sketched in Fig. 3(d~f), which naturally accounts for the long-range positional correlations as appearing in RDF and projected RDFs and the high atomic mobility as revealed in self-diffusivity. For this state to occur, it requires anisotropy in the potential energy landscape to facilitate anisotropic particle diffusion. Anisotropy is rare in simple elemental metals, but gradually becomes common with increased compression,[23] which is partially ascribed to the population of electrons on *p* orbitals that was predicted at both low pressures of 300 GPa[23] and ultra-high pressures of 3 TPa.[24] The excitation of electrons to *p* orbitals in dense hydrogen[24,25] might be the ultimate source of the anisotropy encountered here.

With further increase of the temperature, the *lattice* of this exotic state is gradually dissolved, eventually enters a homogeneous and isotropic liquid. The variation of potential energy and pressure along the isochore for this transition is continuous and smooth. This continuous feature also reflects in the change of proton density distribution, from which the melting point can roughly be estimated at between 1100 and 1300 K for the 2KPM classical model. The mean field theory dictates that any 3D melting should be first order.[3] In this system, we did not find strong hysteresis around the melting; therefore it is likely a weak first-order transition. Alternatively, it also could be a physical realization of the continuous melting (or critical melting), which has been hypothesized and pursued for a long time.[26-28]

In summary, we present a clear and intuitive example in which liquid mobility and crystalline long-range positional ordering coexist harmonically in both quantum and classical regimes, and demonstrate unequivocally that the exotic state observed in dense hydrogen might be a mobile solid. The solid-liquid duality in this state strongly implies a *supersolid* state when protons become quantum, since the fluid component of dense hydrogen was predicted to undergo a superfluid transition.[29] This mechanism is different from that proposed by Andreev and Lifshitz, where the delocalization of boson defectons is the key to make $^4$He have macro-scale mobility at ambient pressure.[30] This observation marks the importance to search for a quantum liquid state of dense hydrogen and the emerging of an exotic supersolid state under extreme compressions.

**Methods.** We use VASP[31] to perform the AIMD simulations, with PAW pseudopotential and PBE exchange-correlation functional being employed. The Nose-Hoover thermostat is employed to equilibrate the temperature. The supercell with periodic boundary conditions contains 480 hydrogen atoms. Using such a large supercell reduces the possible finite size effects to a minimal level.[15,24] The energy cutoff for the plane-wave function basis set is 600 eV. The integration over the first Brillouin zone was carried out using a 2×2×2 *k*-point grid. Denser grids of 3×3×3 and 4×4×4 were also used to check the convergence, with the errors in pressure and energy are less than 0.7 GPa and 8 *m*eV/H, respectively. The influence on atomic structure is negligible. The coarse potential model system was generated with a similar method, but use a coarse *k*-points set, *i.e.*, the 2KPM as reported in Ref. (15). This produces a coarse effective inter-atomic potential between hydrogen atoms. The *P2$_1$/c* structure is obtained by melting the *Fddd* phase of dense H into liquid, and then followed by a long time annealing and structure optimization procedure within 2KPM. It is the crystalline phase that is mostly proximate to the mobile solid. It is necessary to point out that previous studies revealed that PBE functional is not accurate enough to describe H$_2$ dissociation quantitatively.[17,18] But its performance in atomic hydrogen beyond 600 GPa actually is very good, as Refs. (32,33) reported. On the other hand, the dissociation of $H_2$ makes the shortest interatomic distance in dense hydrogen at TPa scale slightly larger than in the low-pressure molecular phases, and this counterintuitive phenomenon validates to use the standard pseudopotential of H at the pressures we studied.[24]

The quantum effects of protons are fully included via the path integral formalism of quantum statistical mechanics.[34,35] AI-PIMD simulations were performed using a homemade path integral MD code,[34] in which VASP was employed to produce the required inter-atomic forces. Both 32 and 64 beads along the imaginary time were used to discretize the Trotter decomposition of the propagators to check the convergence quality of the path integral. Though trial simulations showed that 32 beads are enough to reproduce an acceptable path integral when at 50 K, only the highly converged results using 64 beads are reported here. The functions of RDF $g(r)$ and MSD $\langle r^2(t) \rangle = \langle |r_i(t) - r_i(0)|^2 \rangle$ were calculated by ensemble average over the centroid coordinates of each particle. The self-diffusion constant $D$ is calculated as the long time limit of the slope of the MSD, by using the Einstein relation $D = \frac{1}{6} \lim_{t \to \infty} \frac{\langle r^2(t) \rangle}{t}$. The proton density distribution is generated by ensemble statistics of AIMD and AI-PIMD trajectories over a time range of 2 ps, by using the formula of $\rho(\boldsymbol{r}) = \frac{1}{N} \langle \sum_{i=1}^{N} \delta(\boldsymbol{r} - \boldsymbol{r}_i) \rangle$. The initial configuration of the mobile solid in 2KPM model is generated from the inter-mediate phase in the melting process of *Fddd* phase.[15] Additional thermal cycle processes, *i.e.*, heating and followed carefully annealing, were carried out using *NPT* ensemble to ensure its thermodynamic stability and to remove all possible dislocations and defects. In the classical case with a coarse potential model (2KPM), a perfect crystalline phase *$P2_1/c$* was obtained. Full AIMD and AI-PIMD simulations were then carried out with canonical *NVT* ensemble starting from this structure, which spontaneously becomes a mobile solid when NQE is included or when the temperature is above 350 K in the classical 2KPM case. This procedure guarantees that the obtained mobile solid state is thermodynamically stable. Other settings of the computation and analysis methods are referenced to Ref. (15).

**Acknowledgement.** This work was supported by the National Natural Science Foundation of China under Grant Nos. 11274281 and 11672274, the CAEP Research Projects 2012A0101001 and 2015B0101005. Part of the computation was performed using CCMS of the Institute for Materials Research at Tohoku University, Japan. H.Y.G. appreciates Prof. Roald Hoffmann of Cornell University for reading the manuscript and giving invaluable comments.

**Supporting Information.** Extended discussions on the anisotropy in the mobile solid state, the NQE influence on the melting temperature, the finite size effect, and particle diffusion trajectory.

Author Contributions

H.Y.G. designed and carried out the research. H.Y.G., Q.W. and Y.S. analyzed the data and wrote the manuscript. All authors reviewed the manuscript.

Competing financial interests

The authors declare no competing financial interests.

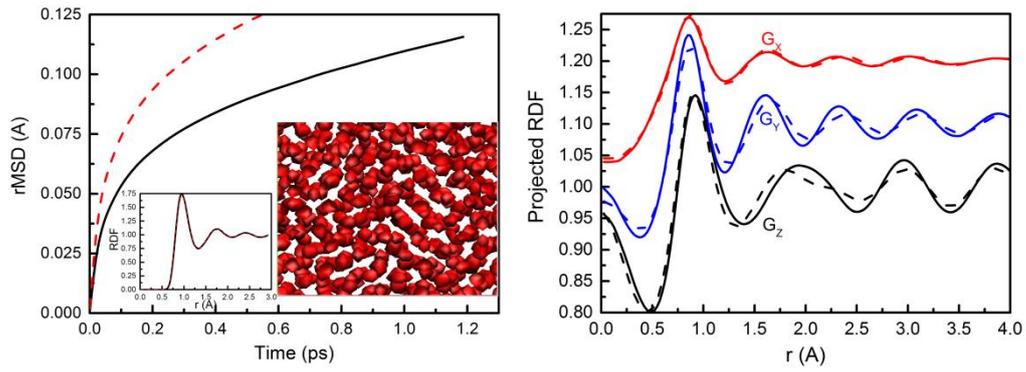

**Figure 1 | An anisotropic non-solid non-liquid state of dense hydrogen.** Left: root mean square displacement (rMSD) of proton centroids and averaged proton density distribution pattern, inset shows the RDF of protons; right: projected RDFs along three Cartesian axes. The solid lines are for the simulations performed at 50 K and dashed lines for at 100 K. The RDFs of 50 and 100 K are almost identical, with small distinctions present in projected RDFs. Note the striking mobility and diffusivity of protons revealed by the rMSD, in contrast to the non-liquid and anisotropic structure uncovered in the proton density distribution and projected RDFs. All results are computed using AI-PIMD with 64 beads at 1.5 TPa of dense hydrogen.

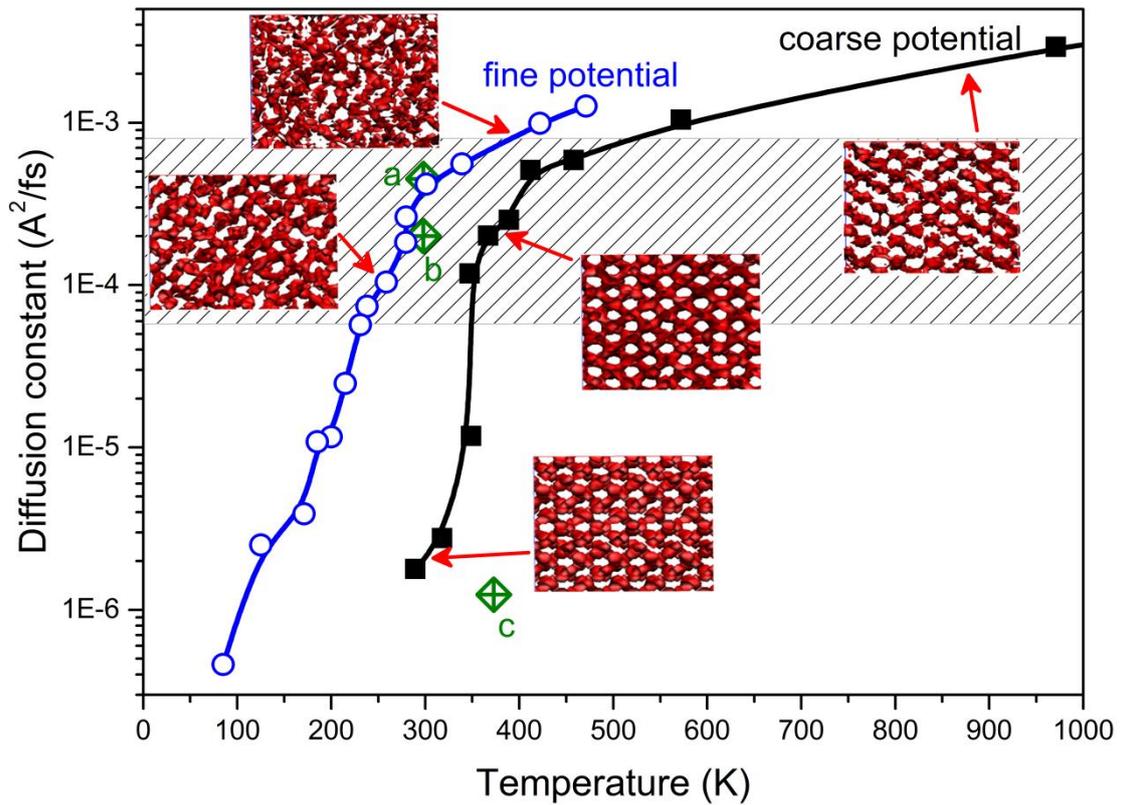

**Figure 2 | High diffusivity and perfect "lattice" observed simultaneously in the exotic state of dense hydrogen.** Proton diffusion constant as a function of temperature for dense hydrogen at about 1 TPa, by comparing the results obtained using the fine potential with those using a degraded coarse potential (2KPM). The typical scale of self-diffusivity in a liquid is indicated by the hatched band, where point *a* and *b* mark the diffusivity of air and hydrogen gas in water, and point *c* corresponds the hydrogen in solid iron, respectively.[16] The proton density distribution patterns at the relevant temperatures are also illustrated. Calculations are performed with AIMD.

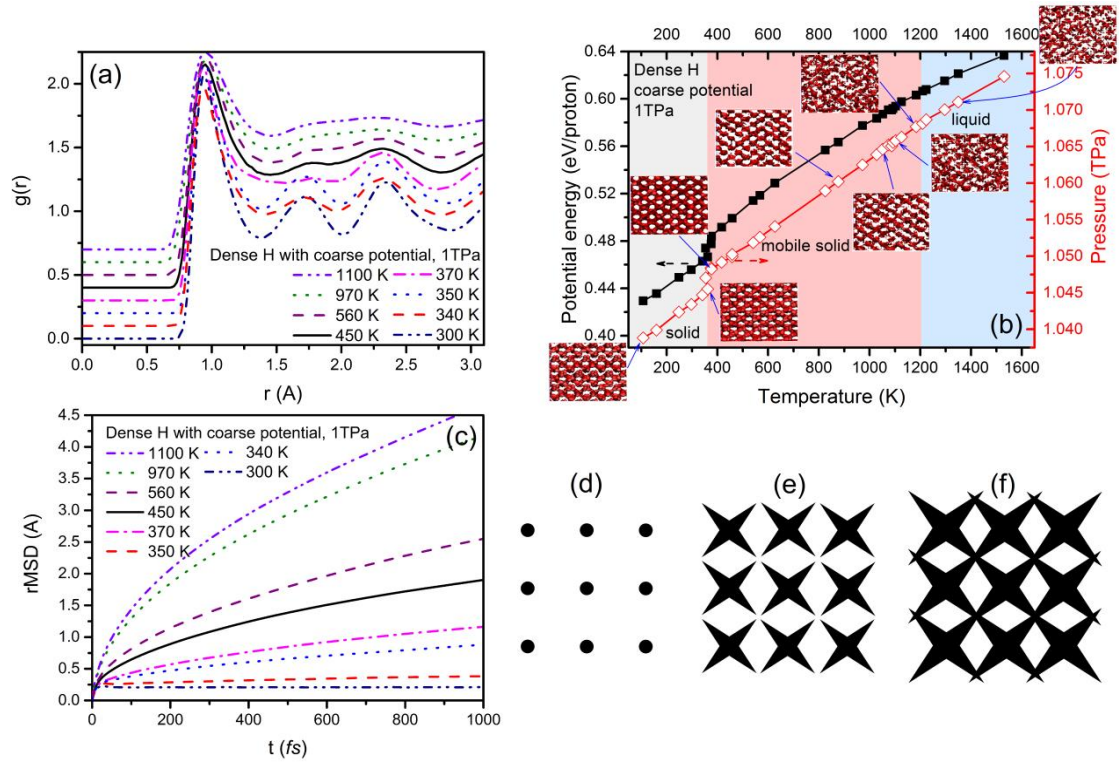

**Figure 3 | A mobile solid state in a physical model of dense hydrogen at 1 TPa.** It is obtained with a coarse potential by slightly degrading the *k*-point sampling quality in the first Brillouin zone (2KPM). This approximation provides a clear-cutting picture of the novel matter state, whereas keeps the underlying physics unchanged. Both RDF (a) and rMSD (c) show typical solid features when temperature below 350 K. A liquid-like RDF is obtained when above 400 K, with great mobility unveiled in the rMSD. This transition also leads to a large jump in potential energy and a discontinuity in the *P-T* plane, as shown in (b). Nevertheless, the positional ordering is still clearly preserved, which is confirmed by the proton density distributions. This phase must be stabilized by certain local anisotropic interactions, with correlated and locally directional motion of protons results in simultaneously flowing and (time averaged) positional ordering. The mechanism of this exotic state developed from a crystalline state is illustrated in terms of particle density distribution patterns in (d)-(f), whereas for an isotropic liquid the shadow region must cover the whole space since the particle density distribution function becomes a constant there. All calculations are carried out with AIMD simulations.

# *Supporting Information*

## 1. Distinction from the Phase IV

It is interesting to note that proton transfer within the graphene-like layers of the phase IV of the molecular dense hydrogen had also been proposed,[1] which could be interpreted as a consequence of the rotation and rebonding of the $H_6$ rings and the hopping of protons among these neighboring motifs.[2] The process is thus more close to the defect migration rather than the bulk diffusion in a fluid. Furthermore, careful studies carried out recently on the structure and dynamics of this phase did not support the previous notion that it has long-range proton transfer or tunneling.[3] Therefore it cannot be considered as a candidate of mobile solid.

## 2. Projected Mean Square Displacement

Similar to decomposition of RDF into the projections along the Cartesian directions,[4] the diffusivity of atoms can also be projected onto orthogonal components. According to the definition, the mean square displacement (MSD) is given by

$$\text{MSD}(\Delta t) = \langle |\boldsymbol{r}(t + \Delta t) - \boldsymbol{r}(t)|^2 \rangle$$

where angle brackets indicate ensemble average. Its projection onto Cartesian direction $\alpha$ is

$$\text{MSD}_\alpha(\Delta t) = \langle |\boldsymbol{r}_\alpha(t + \Delta t) - \boldsymbol{r}_\alpha(t)|^2 \rangle, \qquad \alpha = \text{X, Y, or Z}.$$

The projected root mean square displacement (rMSD) thus reads $\text{rMSD}_\alpha(t) = \sqrt{\text{MSD}_\alpha(t)}$, which can reveal the anisotropic diffusion of atoms. As shown in the right panel of Figure S1, a clear ballistic to diffusive transition is unveiled by projected rMSDs, which is obscured in the total rMSD. This short-time ballistic regime suggests a flat basin in the potential well. Furthermore, the divergent diffusion behavior along different projection directions reflects the anisotropic feature of the basins and the difficulty (or the averaged energy barrier) when particles drift through different diffusion channels.

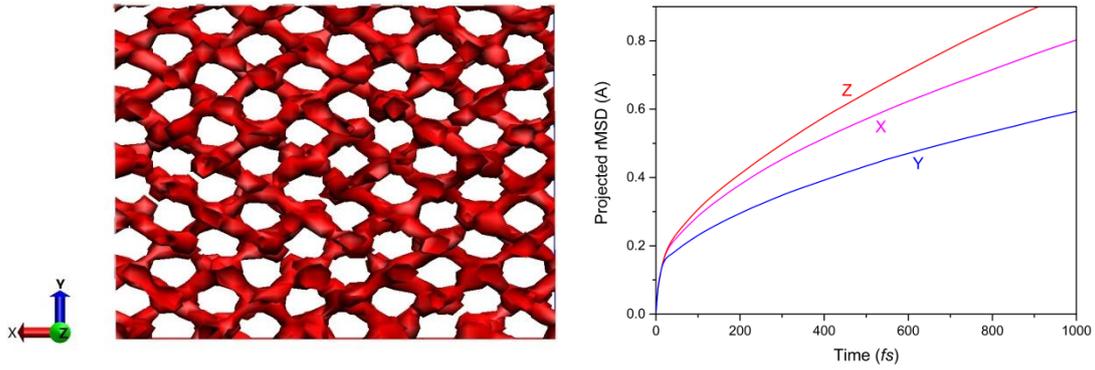

**Figure S1.** Anisotropic diffusivity of classical protons at 390 K and ~1 TPa, calculated with AIMD and corresponds to the coarse potential model as shown in Figures 2 and 3 of the main text. Left: long-time averaged proton density distribution, which unveils a regular "lattice" on the XY plane and the migration pathways that connect the "lattice sites" (Note: the whole "lattice" is actually a 3D network), and right: the projected rMSD that shows fast diffusion along Z and X directions, and sluggish mobility along Y direction. Note that they are consistent with the projected RDFs as shown in Figures 1 and S4, where the strong spatial ordering presents in the plane (*e.g.*, the XY plane of $G_Z$) perpendicular to the projection direction (*e.g.*, the Z direction of $G_Z$).

In principle, rMSD conveys the same information as MSD. Nonetheless, in some situations MSD may be more intuitive and convenient, because of its direct connection to the diffusion constant by the Einstein relation $D = \frac{1}{6}\lim_{t\to\infty}\frac{\langle r^2(t) \rangle}{t}$. For this reason, the MSD corresponding to the Figure 3c in the main text is also given below for reference.

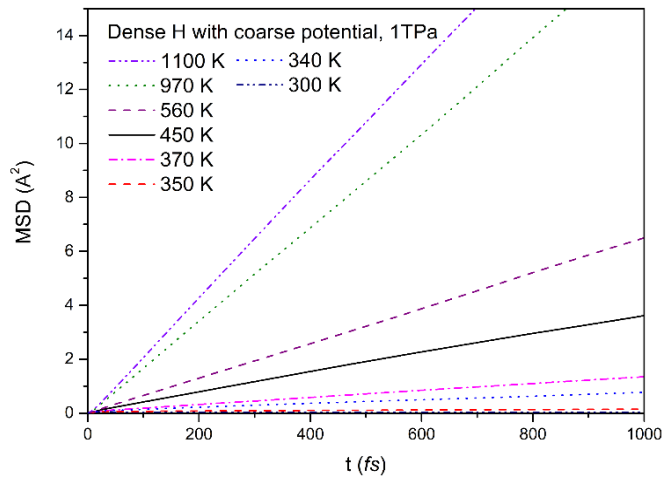

**Figure S2.** Mean square displacement (MSD) of the 2KPM model, corresponding to that of Figure 3c in the main text.

## 3. Particle Density Distribution of Typical Solid

To give an intuitive impression of how drastically the hydrogen atoms in the mobile solid phase drift, the averaged particle density distribution of gold, as a typical metallic solid, is shown in Figure S3. It can be seen that even when nearing the melting temperature, the particles in this typical metallic solid are still well localized to their positions, in a sharp contrast against what is shown in Figure S1.

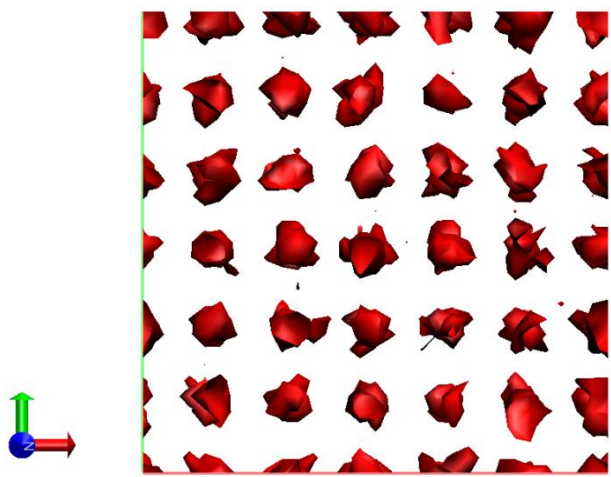

**Figure S3.** Long-time averaged atomic density distribution of gold when approaching (but still lower than) the melting temperature from the FCC solid phase. The motion of each individual atom is restricted to well-separated pockets (a typical property for a normal solid), which is sharply in contrast to that shown in Figure S1 and those displayed in the main text.

## 4. Projected Pair Correlation Function

The radial distribution function (RDF) can be decomposed into projected RDFs by using the projection technique introduced in ref 4. The results for the 2KPM model calculated with AI-PIMD using 8 beads, which has taken the nuclear quantum effects (NQE) into account, are shown in Figure S4. It not only reveals the anisotropy in this system, but also gives an estimation of the melting temperature of this mobile solid. Its value of 950 K is lower than the classical case (as suggested in the Figure 3 of the main text) where NQE has been excluded, and is consistent with the general expectation that zero-point motion (or generally the NQE) prefers isotropic states, thus would lower the transition temperature.

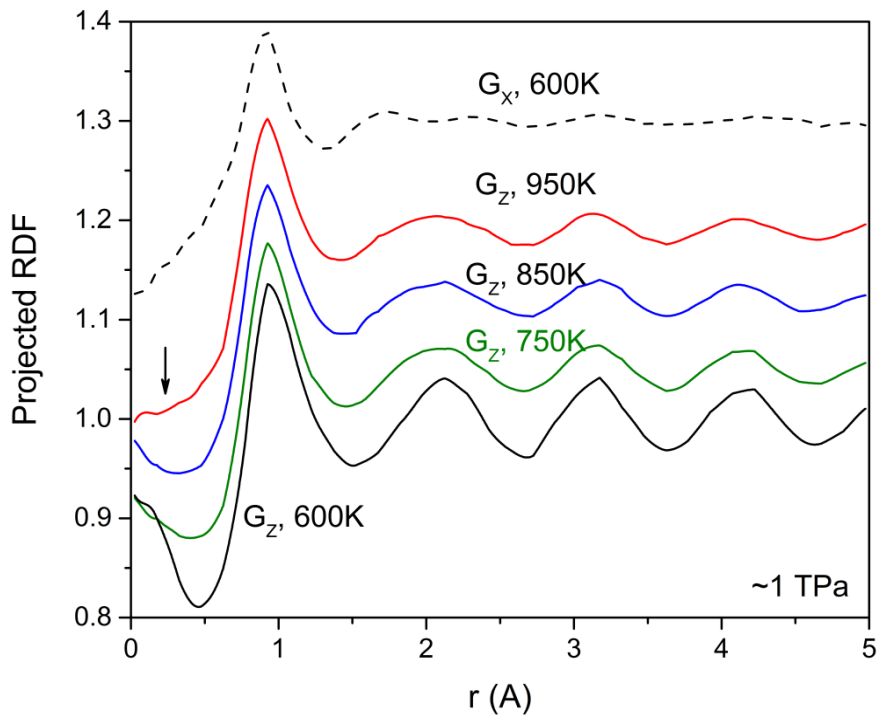

**Figure S4.** Variation of the projected RDF $G_Z(r)$ with temperature at a pressure of ~1 TPa calculated with AI-PIMD and corresponds to the coarse potential model (2KPM) as shown in Figures 2 and 3 of the main text. Eight beads along the imaginary time were used to discretize the Trotter decomposition of the propagators in the path integral. The $G_X(r)$ at 600 K is plotted for comparison. The arrow indicates the beginning of the homogeneous liquid feature on these projected RDFs. The gradual losing of the anisotropy is apparent, implying a possible continuous melting of this exotic state. Since this calculation includes NQE, the classical melting temperature of this model system is higher than 950 K.

## 5. Finite Size Effect

For MD simulations of dense hydrogen, a lot of previous works have demonstrated that a supercell with a size containing more than 200 hydrogen atoms is large enough to reach converged thermodynamics with negligible finite size effects.[4] Such cell size is also adequate for modelling the melting curve.[4,5] In simulation of the mobile solid phase, we used a supercell with 480H. It should be large enough to eliminate any possible finite size effect. In order to confirm this argument, we also carried out a simulation with a lager supercell containing 960H. The obtained result (with the corresponding proton density distribution at 450 K displayed in Figure S5) is very similar to that of the supercell having 480H, demonstrating that the properties of mobile solid phase have been correctly captured. We also noticed that for some cell size dimensions, the orientation of the *lattice* ordering might slightly change. This is explainable since the phase is flexible, and should be able to adjust itself accordingly to match the boundary conditions by self-organization of the mobile particles.

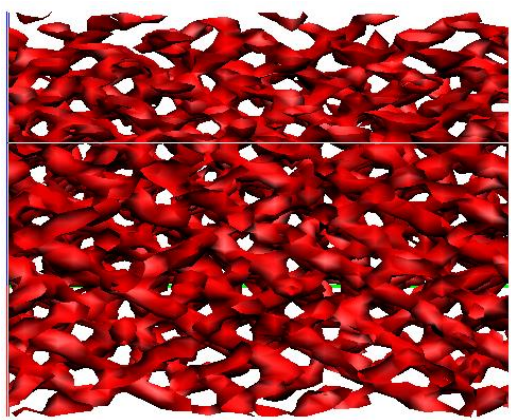

**Figure S5.** Particle density distribution of the 2KPM model in a large supercell containing 960H at 450 K. The mobile solid characteristics can be clearly seen.

## 6. Particle Diffusion Trajectory

To illustrate the mobility character of particles in the mobile solid state, the diffusion trajectory of two nearest neighboring particles in the 2KPM model at 500 K is plotted in Figure S6. It can be seen that though these two particles are initially as the nearest neighbors, there is no obvious correlation in their diffusion path. One of them hops back and forth mainly between two "lattice sites", whereas the other one drifts far away and traverses several "lattice sites". However, all diffusion pathways must be correlated and concerted in the sense of long-time statistics, and it is these "lattice sites" and the particle diffusion pathways that form the ordered pattern in the mobile solid phase, which is an evident self-organizing phenomenon as shown in Figure 3 of the main text.

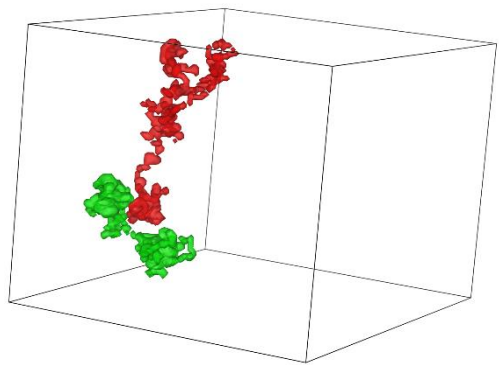

**Figure S6.** Diffusion trajectory of two nearest neighboring particles in the 2KPM model at 500 K.

## ■ SUPPORTING REFERENCES